\documentclass[twocolumn,showpacs,preprintnumbers,aps]{revtex4}

\newcommand{\BE}{\begin{equation}}
\newcommand{\EE}{\end{equation}}
\newcommand{\BA}{\begin{eqnarray}}
\newcommand{\EA}{\end{eqnarray}}
\usepackage{graphicx}
\usepackage{dcolumn}
\usepackage{bm}
\input epsf.tex

\begin{document}

\title{A simulation study of localization of electromagnetic waves in two-dimensional random dipolar systems}

\author{Ken Kang-Hsin Wang\footnote{Previously publishing as Kang-Xin Wang}} \author{Zhen
Ye}\email{zhen@phy.ncu.edu.tw} \affiliation{Wave Phenomena
Laboratory, Department of Physics, National Central University,
Chungli, Taiwan 32054}

\date{February 25, 2003}

\begin{abstract}

We study the propagation and scattering of electromagnetic waves
by random arrays of dipolar cylinders in a uniform medium. A set
of self-consistent equations, incorporating all orders of multiple
scattering of the electromagnetic waves, is derived from first
principles and then solved numerically for electromagnetic fields.
For certain ranges of frequencies, spatially localized
electromagnetic waves appear in such a simple but realistic
disordered system. Dependence of localization on the frequency,
radiation damping, and filling factor is shown. The spatial
behavior of the total, coherent and diffusive waves is explored in
detailed, and found to be comply with a physical intuitive
picture. A phase diagram characterizing localization is presented,
in agreement with previous investigations on other systems.

\end{abstract}

 \maketitle

\section{Introduction}

The concept of localization was originally introduced by
Anderson\cite{Anderson} for electrons in a crystal. In the case of
a perfectly periodic lattice, except in the gaps all the
electronic states are extended and are represented by Bloch
states. When a sufficient amount of disorders is added to the
lattice, for example in the form of random potentials, the
electrons may become spatially localized due to the multiple
scattering by the disorders. In such a case, the eigenstates are
exponentially confined in the space\cite{Lee}. The inception of
the localization concept has opened a new era for the study of
electrons in disordered systems, and stimulated a tremendous
research. The concept of localization has also rendered a great
development in many other fields such as seismology\cite{Seis},
oceanology\cite{ocean}, and random lasers\cite{Laser}, to name
just a few. The great efforts have been summarized in a number of
excellent reviews (e. g.
\cite{Lee,Thouless,John,Sheng,Lagen,Rossum,Van,loc}).

Over the past two decades, localization of classical waves has
been under intensive investigations, leading to a very large body
of literature(e.~g.
\cite{John,Sheng,Lagen,HW,Kirk1,He,Condat,Sor,Genack,McCall,McCall2,Marian,Sigalas,AAA,Ye1,Wiersma,weak,AAC}).
Such a localization phenomenon has been characterized by two
levels. One is the weak localization associated with the enhanced
backscattering. That is, waves which propagate in the two opposite
directions along a loop will obtain the same phase and interfere
constructively at the emission site, thus enhancing the
backscattering. The second is the strong localization, without
confusion often just termed as localization, in which a
significant inhibition of transmission appears and the energy is
mostly confined spatially in the vicinity of the emission site.

While the weak localization, regarded as a precursor to the strong
localization, has been well studied both theoretically (e.~g. the
monograph\cite{Sheng}) and experimentally (e.~g. \cite{weak}),
observation of strong localization of classical waves for higher
than one dimension remains a subject of
debate\cite{Wiersma,AAC,debate}, primarily because a suitable
system is hard to find and the observation is often obscured by
such effects as the residual absorption\cite{debate} and
scattering attenuation. Asides from few exceptions\cite{McCall},
most experiments are based on observations of the exponential
decay of waves as they propagate {\it through} disordered media.
This was pointed out in Ref.~\cite{AAC}. According to
\cite{Chen2002}, this type of experiments is {\it not} sufficient
to discern whether the medium really only has localized states.
Unwanted effects of non-localization origin can also contribute to
the exponential decay, making data interpretation possibly
ambiguous. In a conclusion, Sigalas {\it et al.} \cite{Sigalas}
pointed out that there is no conclusive experimental evidence for
localization of electromagnetic waves (EM) in two dimensions (2D).
We mention that there was a report of the observation of microwave
localization in two dimensions when a transmitting source is
inside disordered media\cite{McCall}.

The difficulties in observing localization of EM waves mainly lie
in a couple of problems. First, wave localization only appears for
strongly scattering media, and such a medium is often hard to
find. Second, localization effects are often entangled with other
effects such as dissipation, wave deflection, or boundary
effects\cite{loc}, making data interpretation often ambiguous. In
a recent communication, a simple but seemingly realistic model
system has been proposed to study EM localization in 2D random
media\cite{YLS}. This model originated from the previous study of
the radiative effects of the electric dipoles embedded in
structured cavities\cite{Erdogan}. It was shown that EM
localization is possible in such a disordered system. When
localization occurs, a coherent behavior appears and is revealed
as a unique property differentiating localization from either the
residual absorption or the attenuation effects.

With the present paper, we wish to explore further the
localization properties of the system outlined in \cite{YLS}. We
will investigate the dependence of localization behavior on a
numbers of parameters including frequency, filling factor,
scattering strength, damping effect, and two different ways of
measuring localization. Additionally, the spatial behaviors of the
total, coherent and diffusive wave intensities will be studied,
and are shown to comply with a simple physical intuition.

\section{The system and theoretical formulation}

\subsection{The system}

\begin{figure}[hbt]
\vspace{10pt} \epsfxsize=2.5in\epsffile{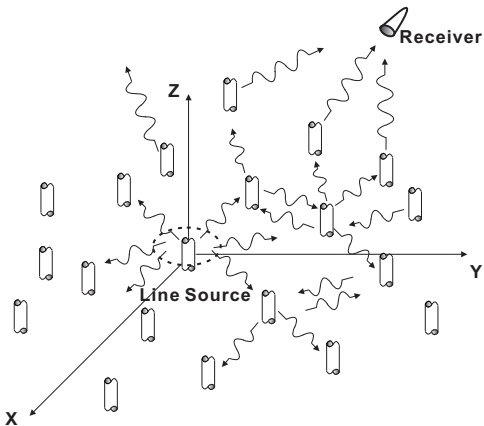}
\caption{Conceptual layout of the system and the simulation.}
\label{fig1}
\end{figure}

Following Erdogan et al.\cite{Erdogan}, we consider 2D dipoles as
an ensemble of harmonically bound charge elements. In this way,
each 2D dipole is regarded as a single dipole line, characterized
by the charge and dipole moment per unit length. Assume that $N$
parallel dipole lines, aligned along the $z$-axis, are embedded in
a uniform dielectric medium and {\it randomly} located at
$\vec{r}_i (i=1,2,\dots,N)$. The averaged distance between dipoles
is $d$. A stimulating dipole line source is located at
$\vec{r}_s$, transmitting a continuous wave of angular frequency
$\omega$. By the geometrical symmetry of the system, we only need
to consider the $z$ component of the electrical waves. The
conceptual layout of the system is illustrated by Fig.~\ref{fig1}.

Here it is shown that a line source is located at the origin, and
the receiver is located at some distance from the source. After
multiply scattered, the transmitted waves reach at the receiver.

\subsection{The formulation}

Although much of the following materials can be referred to in
\cite{YLS}, we repeat the important parts here for the sake of
convenience and completeness.

Upon stimulation, each dipole will radiate EM waves. The radiated
waves will then repeatedly interact with the dipoles, forming a
process of multiple scattering. The equation of motion for the
$i$-th dipole is \BA \frac{d^2}{dt^2}p_i + \omega_{0,i}^2p_i &=&
\frac{q_i^2}{m_i}E_{z}(\vec{r}_i) - b_{0,i}\frac{d}{dt}
p_i,\nonumber\\
& & \mbox{for} \ i = 1, 2,\dots, N. \label{eq:1}\EA where
$\omega_{0,i}$ is the resonance (natural) frequency, $p_i$, $q_i$
and $m_i$ the dipole moment, charge and effective mass per unit
length of the $i$-th dipole respectively. $E_{z}({\vec{r}_i})$ is
the total electrical field acting on dipole $p_i$, which includes
the radiated field from other dipoles and also the directly field
from the source. The factor $b_{0,i}$ denotes the damping due to
energy loss and radiation, and can be determined by energy
conservation. Without energy loss (to such as heat), $b_{0,i}$ can
be determined from the balance between the radiative and
vibrational energies, and is given as\cite{Erdogan} \BE b_{0,i} =
\frac{q_i^2\omega_{0,i}}{4\epsilon m_i c^2}, \label{eq:2}\EE with
$\epsilon$ being the constant permittivity and $c$ the speed of
light in the medium separately.

Equation (\ref{eq:1}) is virtually the same as Eq.~(1) in
\cite{Erdogan}. The only difference is that in \cite{Erdogan},
$E_z$ is the reflected field at the dipole due to the presence of
reflecting surrounding structures, while in the present case the
field is from the stimulating source and the radiation from all
other dipoles.

The transmitted electrical field from the continuous line source
is determined by the Maxwell equations\cite{Erdogan}\BE \left
(\nabla^2 - \frac{\partial^2}{c^2\partial t^2}\right
)G_0(\vec{r}-\vec{r}_s) =
-4\mu_0\omega^2p_0\pi\delta^{(2)}(\vec{r}-\vec{r}_s) e^{-i\omega
t}, \label{eq:3}\EE where $\omega$ is the radiation frequency, and
$p_0$ is the source strength and is set to be unit. The solution
of Eq.~(\ref{eq:3}) is clearly \BE G_0(\vec{r}-\vec{r_s}) =
(\mu_0\omega^2p_0) i\pi H_0^{(1)}(k|\vec{r}-\vec{r}_s|)
e^{-i\omega t}, \label{eq:4}\EE with $k=\omega/c$, and $H_0^{(1)}$
being the zero-th order Hankel function of the first kind.

Similarly, the radiated field from the $i$-th dipole is given by
\BE \left (\nabla^2 - \frac{\partial^2}{c^2\partial t^2}\right
)G_i(\vec{r}-\vec{r}_i) =
\mu_0\frac{d^2}{dt^2}p_i\delta^{(2)}(\vec{r}-\vec{r}_i).
\label{eq:5}\EE The field arriving at the $i$-th dipole is
composed of the direct field from the source and the radiation
from all other dipoles, and thus is given as \BE E_z(\vec{r}_i) =
G_0(\vec{r}_i - \vec{r}_s) + \sum_{j=1, j\neq i}^N G_j(\vec{r}_i -
\vec{r}_j). \label{eq:6}\EE

Substituting Eqs.~(\ref{eq:4}), (\ref{eq:5}), and (\ref{eq:6})
into Eq.~(\ref{eq:1}), and writing $p_i = p_ie^{-i\omega t}$, we
arrive at \BA (-\omega^2 + \omega^2_{0,i} - i\omega b_{0,i})p_i =&
&\nonumber\\ \frac{q_i^2}{m_i}\left[G_0(\vec{r}_i - \vec{r}_s) +
\sum_{j=1, j\neq i}^N \frac{\mu_0 \omega^2}{4}
iH_0^{(1)}(k|\vec{r}_i-\vec{r}_j|)p_i\right].& &  \label{eq:7}\EA
This set of linear equations can be solved numerically for $p_i$.
After $p_i$ are obtained, the total field at any space point can
be readily calculated from \BE E_z(\vec{r}) = G_0(\vec{r} -
\vec{r}_s) + \sum_{j=1}^N G_j(\vec{r} - \vec{r}_j).\label{eq:8}\EE

In the standard approach to wave localization, waves are said to
be localized when the square modulus of the field
$|E(\vec{r})|^2$, representing the wave energy, is spatially
localized after the trivial cylindrically spreading effect is
eliminated. Obviously, this is equivalent to say that the further
away is the dipole from the source, the smaller its oscillation
amplitude, expected to follow an exponentially decreasing pattern.

To this end, it is instructive to point out that an alternative
two dimensional dipole model was devised previously by Rusek and
Orlowski\cite{Marian}. The authors derived a set of linear
algebraic equations, which is similar in form to the above
Eq.~(\ref{eq:6}). However, there are some fundamental
discrepancies between the two models. In \cite{Marian}, the
interaction between dipoles and the external field is derived by
the energy conservation, while in the present case the coupling is
determined without ambiguity by the Newton's second law. The
former leads to an undetermined phase factor. According to, e.~g.
Refs.~\cite{Erdogan,ccw}, the energy conservation can only give
the radiation factor in Eq.~(\ref{eq:2}). We would also like to
point out that the set of couple equations in Eq.~(\ref{eq:7}) is
similar in spirit to the tight-binding model used to study the
electronic localization\cite{Anderson,Zallen}.

There are several ways to introduce randomness to
Eq.~(\ref{eq:7}). For example, the disorder may be introduced by
randomly varying such properties of individual dipoles as the
charge, the mass or the two combined. This is the most common way
that the disorder is introduced into the tight-binding model for
electronic waves\cite{Zallen}. In the present study, the disorder
is brought in by the random distribution of the dipoles.

For simplicity yet without losing generality, assume that all the
dipoles are identical and they are randomly distributed within a
square area. The source is located at the center (set to be the
origin) of this area. For convenience, we make Eq.~(\ref{eq:7})
non-dimensional by scaling the frequency by the resonance
frequency of a single dipole $\omega_0$. This will lead to two
independent non-dimensional parameters $b = \frac{q^2\mu_0}{4m}$
and $b_0^\prime =
\frac{\omega}{\omega_0}\left(\frac{b_0}{\omega_0}\right)$. Both
parameters may be adjusted in the experiment. For example, the
factor $b_0$ can be modified by coating layered structures around
the dipoles\cite{Erdogan}. Then Eq.~(\ref{eq:7}) becomes simply
\BA (-f^2+1 -ib_0^\prime)p_i =
& & \nonumber\\
ibf^2\left[p_0H_0^{(1)}(k|\vec{r}_i-\vec{r}_s|) + \sum_{j=1, j\neq
i}^N p_iH_0^{(1)}(k|\vec{r}_i-\vec{r}_j|)\right] && \label{eq:10})
\EA with $f=\frac{\omega}{\omega_0}$. Eq.~(\ref{eq:10}) is
self-consistent and can be solved numerically for $p_i$ and then
the total field is obtained through Eqs.~(\ref{eq:3}),
(\ref{eq:5}) and (\ref{eq:8}).

\section{The results and discussion}

\subsection{Two numerical measuring scenarios}

\begin{figure}[hbt]
\vspace{10pt} \epsfxsize=2.5in\epsffile{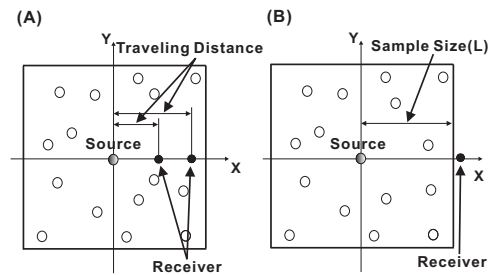} \caption{The
conceptual layout of the two measuring methods.} \label{fig2}
\end{figure}

In the following computation, we will consider two scenarios. They
are illustrated in Fig.~\ref{fig2} with the coordinates being
shown. In both cases, the sample takes a fixed rectangular shape
of which the size may vary. The dipoles, denoted by the small
circles, are placed within the rectangle in a complete randomness.
The receiver, denoted by the filled black circle, is placed on the
$x$-axis.

In case (A), the sample size is fixed. The receiver is placed
along $x$-axis to measure the signal at various positions,
yielding the result of the transmission signal versus the distance
between the source and the receiver, termed as the travelling
distance in the paper. This scenario complies with the original
conjecture of observation of localization\cite{Lee}. In case (B),
the receiver is placed at a very small distance outside the
sample. While the sample size varies, this method is to measure
the transmission through the sample of various sizes, yielding the
result of the transmission signal versus the sample size.

\subsection{General discussion of localization}

Before moving to solve Eq.~(\ref{eq:7}) for the phenomenon of
localization of EM waves, we discuss some general properties of
wave localization.

{\it The coherence in localization} Although some parts of the
following discussion has been reported earlier, we repeat here for
the sake of convenience and importance. The energy flow of EM
waves is $\vec{J} \sim \vec{E}\times\vec{H}$. By invoking the
Maxwell equations to relate the electrical and magnetic fields, we
can derive that the time averaged energy flow is \BE <\vec{J}>_t
\equiv \frac{1}{T}\int_0^T dt \vec{J} \sim
|\vec{E}|^2\nabla\theta, \label{eq:9}\EE where the electrical
field is written as $\vec{E} = \vec{e}_E |\vec{E}|e^{i\theta}$,
with $\vec{e}_E$ denoting the direction, $|\vec{E}|$ and $\theta$
being the amplitude and the phase respectively. It is clear from
Eq.~(\ref{eq:9}) that when $\theta$ is constant, at least by
spatial domains, while $|\vec{E}| \neq 0$, the flow would come to
a stop and the energy will be localized or stored in the space. In
the localized state, a source can no longer radiate energies.
Alternatively, we can write the oscillation of the dipoles as $p_i
= |p_i|e^{i\theta_i}$. By studying the square modulus of $p_i$ in
the form of $|\vec{r}_i-\vec{r}_s||p_i|^2$, and its phase
$\theta_i$, we can also investigate the localization of EM waves.
Note here that the factor $|\vec{r}_i-\vec{r}_s|$ is to eliminate
the cylindrical spreading effect in 2D as can be seen from the
expansion of the Hankel function $|H_0^{(1)}(x)|^2 \sim
\frac{1}{x}$. That the phase $\theta$ is constant implies that a
coherence behavior appears in the system, i.~e. the localized
state is a phase-coherent state, as previously
discussed\cite{Ye2}. It is a unique feature of wave localization,
and has also been shown to be related to electronic localization
(e.~g. Ref.~\cite{Gurtvitz}).

{\it Spatial behavior of localized waves} Following \cite{3D}, a
general consideration of the spatial behavior of localized waves
is possible. Consider a wave transmitted in a random medium. The
transport equation for the total energy intensity $I$, i.~e.
$<|E|^2>$, may be intuitively written as \BE \frac{dI}{dx} =
-\alpha I, \label{eq:1a} \EE where $\alpha$ represents decay along
the path traversed. After penetrating into the random medium, the
wave will be scattered by random inhomogeneities. As a result, the
wave coherence starts to decrease, yielding the way to
incoherence. Extinction of the coherent intensity $I_C$, i.~e.
$|<E>|^2$, is described by \BE \frac{dI_C}{dx} = -\gamma I_C,
\label{eq:2a} \EE with the attenuation constant $\gamma$.
Eqs.~(\ref{eq:1a}) and (\ref{eq:2a}) lead to the exponential
solutions \BE I(x) = I(0)e^{-\alpha x}, \ \ \ \mbox{and} \ \ \
I_C(x) = I(0)e^{-\gamma x}. \EE In deriving these equations, the
boundary condition was used; it states that $I(0) = I_C(0)$ as no
scattering has been incurred yet at the interface. According to
energy conservation, the incoherent intensity $I_D$ (diffusive) is
thus \BE I_D(x) = I(x) - I_C(x). \EE

When there is no absorption, the decay constant $\alpha$ is
expected to vanish and the total intensity will then be constant
along the propagation path. Then, the coherent energy gradually
decreases due to random scattering and transforms to the diffusive
energy, while the sum of the two forms of energy remains a
constant. This scenario, however, changes when localization
occurs. Even without absorption, the total intensity can be
localized near the interface due to multiple scattering. When this
happens, $\alpha$ does not vanish. The transport of the total
intensity may be still described by Eq.~(\ref{eq:1a}), and the
inverse of $\alpha$ would then refer to the localization length.

\begin{figure}[hbt]
\vspace{10pt} \epsfxsize=2.5in\epsffile{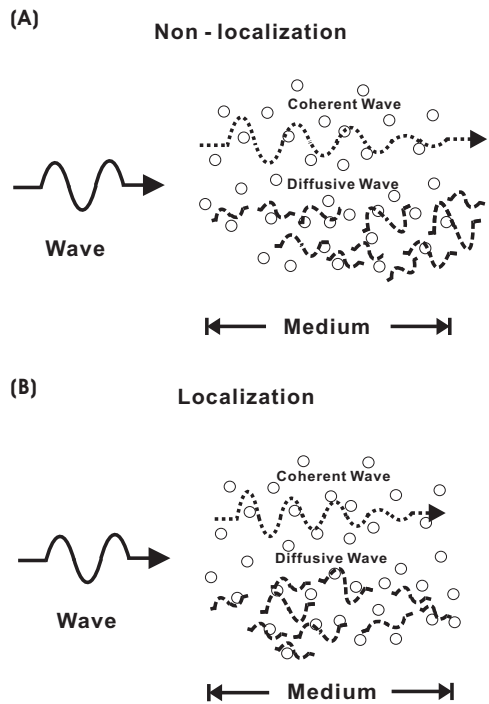}
\caption{Picturised non-localization and localization effects.}
\label{fig3}
\end{figure}

The above perceptual description may be illustrated by
Fig.~\ref{fig3}. Without or with little absorption and with no
localization, the energy propagation is anticipated to follow the
behavior depicted in (A). When the localization comes in function,
the wave will be trapped within an $e$-folding distance from the
penetration, as prescribed by (B). In the non-localization case,
the diffusive intensity increases steadily as more and more
scattering occurs, complying with the Milne diffusion. In the
localized state, the diffusion energy increases initially. It will
be eventually stopped by the interference of multiple scattering
waves. Issues may be raised with respect to whether this
apprehended image is supported by actual situations. In the rest,
we will inspect this problem.

Comparing Fig.~\ref{fig3} (A) and (B), we know that one of the key
differences between non-localization and localization features
lies in the difference in the behavior of the diffusive waves.
However, the eventual task of inferring localization is to find
whether the total energy decays with the sample size.

\subsection{Numerical results}

Unless otherwise noted, the following parameters are used in the
numerical simulation: the non-dimensional damping rate,
$b_0/\omega_0 = 0.001$ and the interaction coupling, $b = 0.001$.
The filling factor ($\beta$) varies from 2.25 to 25; the filling
factor is defined as the number of dipoles per unit area. But
without notification, the filling factor is taken as 6.25. The
number of random configurations for averaging is taken in such a
way that the convergency is assured. In the calculation, we scale
all lengths by a length $D$ such that $k_0D=1$, and frequency by
$\omega_0$. In this way, the frequency always enters as $k/k$. We
find that all the results shown below are only dependent on
parameters $b$, $b_0/\omega_0$, and the ratio $\omega/\omega_0$ or
equivalently $k/k_0$. Such a simple scaling property may
facilitate designing experiments. In the numerical computation, we
take $c=1$ for convenience. The total wave at a spatial point is
scaled as $T(\vec{r}) \equiv E(\vec{r})/E_0(\vec{r})$, with $E_0$
being the direct wave from the source, so that the trivial
geometric spreading effect is naturally removed.

\begin{figure}[hbt]
\vspace{10pt} \epsfxsize=2.75in\epsffile{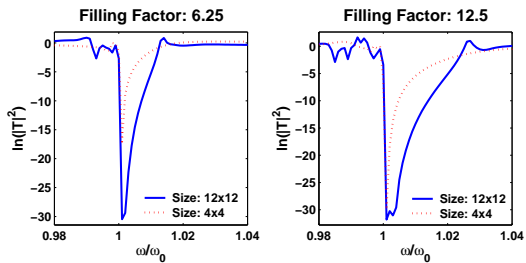}
\caption{Transmission versus frequency for two filling factors and
two sample sizes.} \label{fig4}
\end{figure}

First, we plot the frequency response of the transmission in the
scenario (B) of Fig.~\ref{fig2}. The results are shown in
Fig.~\ref{fig4}. Here we see that there is a narrow window within
which the transmission is highly inhibited, implying a strong
localization effect. It is also clear that when the sample size is
increased, the inhibition increases. Comparing the results for the
two filling factors, we know that the strong inhibition regime
increases with filling factor. In the next computations, we will
focus on frequencies within the strong inhibition region.

\begin{figure}[hbt]
\vspace{10pt} \epsfxsize=3in\epsffile{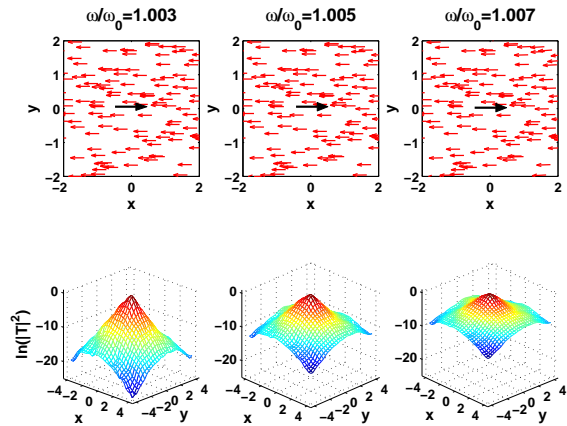} \caption{The phase
diagram for the two dimensional phase vectors defined in the text;
the phase of the source is assumed to be zero. Each vector is
located at the site of the dipole; thus the locations of the phase
vectors also denote the random distribution of the dipoles.
Bottom: The spatial distribution of energy ($\sim |T|^2$). Three
frequencies are chosen, the sample size is $8\times 8$.}
\label{fig5}
\end{figure}

Now we consider the phase and the spatial distribution of energy
of the system. To describe the phase behavior of the system, we
assign a unit phase vector, $\vec{u}_i = \cos\theta_i\vec{e}_x +
\sin\theta_i\vec{e}_y$ to the oscillation phase $\theta_i$ of the
dipoles. Here $\vec{e}_x$ and $\vec{e}_y$ are unit vectors in the
$x$ and $y$ directions respectively. These phase vectors are
represented by a phase diagram in the $x-y$ plane with the phase
vector $\vec{u}_i$ being located at the dipole to which the phase
$\theta_i$ is associated. The results are depicted for three
frequencies by Fig.~\ref{fig5}.

Here we see clearly shown that for the three frequencies within
the strong inhibition region, the energy is spatially confined
near the transmitting source, and, as expected, the energy seems
to decrease nearly exponentially along any radial direction.
Meanwhile, the system reveals an in-phase phenomenon: nearly all
the phase vectors of the dipoles point to the same direction,
exactly opposite to the phase vector of the source which is
denoted by the black arrow. The picture represented by
Fig.~\ref{fig5} fully complies with the general description of the
coherence in localization stated above.

We also note from Fig.~\ref{fig5} that near the sample boundary,
the phase vectors start to point to different directions. This is
because the numerical simulation is carried out for a finite
sample size. For a finite system, the energy can leak out at the
boundary, resulting in disorientation of the phase vectors. When
enlarging the sample size by adding more dipoles while keeping the
averaged distance between dipoles fixed, the area showing the
phase coherence will increase accordingly.

\begin{figure}[hbt]
\vspace{10pt} \epsfxsize=2.75in\epsffile{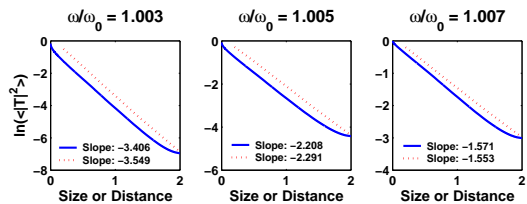}
\caption{Comparison of the total transmission at three frequencies
for the two scenarios shown by Fig.~\ref{fig2}. The x-labels
`Distance' and `Size' refer to scenario (A) and (B) respectively.}
\label{fig6}
\end{figure}

The results of Fig.~\ref{fig5} are encouraging, as they are a
strong indication of localization. In the following, we will
further explore the features of localization. In Fig.~\ref{fig6},
we compare the transmission results for the two scenarios from
Fig.~\ref{fig2}. Here is shown that although there is a slight
difference in the transmitted strength, overall speaking the
spatial decay features are nearly identical, signified by the
match of the decaying slopes indicated in the figure. Though
suspected to be true previously, such a match is important, and to
the best of our knowledge this is the first that has been ever
shown for EM waves. It supports directly that the scenario (B) can
also be used to infer localization effects, facilitating
measurements of localization; in the original conjecture, it was
scenario (A) that has been suggested for discerning localization.
In the rest of simulation, we will adopt scenario (B) in
Fig.~\ref{fig2}.

\begin{figure}[hbt]
\vspace{10pt} \epsfxsize=2.75in\epsffile{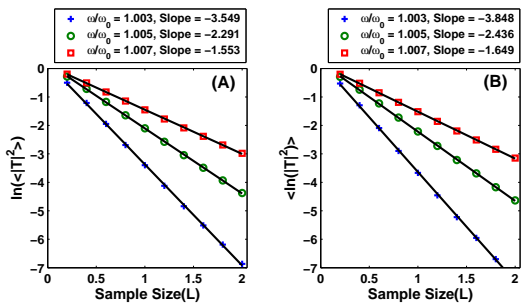} \caption{The
transmission versus the sample size for three frequencies. The
slopes can be used to estimate the localization length.}
\label{fig7}
\end{figure}

The bottom panel of Fig.~\ref{fig6} indicates that the level of
spatial localization of energy varies for the three frequencies.
To quantify the localization in Fig.~\ref{fig6}, we plot the total
energy as a function of the sample size. The results are presented
by Fig.~\ref{fig7}. Here, the numerical data are fitted with the
least squares method and the fitted curves are shown by the solid
lines; the unnoticeable deviation from the lines reflects the
fluctuation due to the random distribution. Two ways of averaging
are adopted. One is the traditional way in which the logged total
transmission is averaged, while the other is to take log of the
averaged total energy. These can be referred to from the y-axis
labels. It shows that after removing the spreading factor, the
data can be fitted by $e^{-r/\xi}$. From the slope of the solid
lines, the localization lengths $\xi$, which are the inverse of
the slopes can be estimated. Here is shown that the decaying
slopes from the two averaging methods are very close, an
encouraging fact. It is also indicated by the decreasing slopes
with frequency that the localization effect decreases as the
frequency increases.

\begin{figure}[hbt]
\vspace{10pt} \epsfxsize=2.75in\epsffile{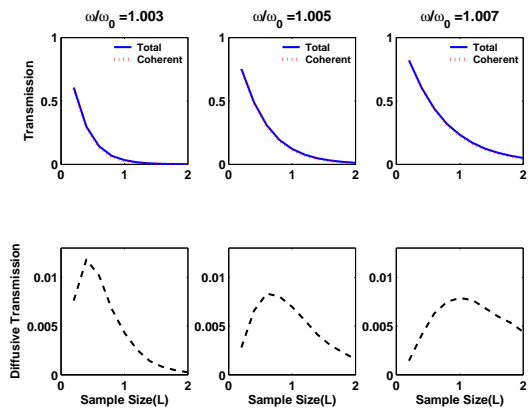}
\caption{Behaviors of the total, coherent, and the diffusive
energies as a function of sample size in the scenario described by
Fig.~\ref{fig2}(B).} \label{fig8}
\end{figure}

With the fixed filling factor of 6.25, we have also investigated
the spatial variations of the total, coherent and the diffusive
energies for three frequencies discussed. The results are
presented in Fig.~\ref{fig8}. Here we see that the results are in
accordance with the previous general consideration of
localization. Due to scattering and localization, the coherent
waves decrease with the sample size. The diffusive wave increases
initially as more and more scattering occurs, then reaches a peak
and starts to decay due to the localization effect, in agreement
with the description of Fig.~\ref{fig3}(B). The results show that
in the present system, the diffusive portion in the total energy
is much smaller than the coherent portion, indicating that the
mean free path is very small. When plotted in the log scale, we
have found that the total and coherent energies decays
exponentially with the distance, while the diffusive energy
increases initially, then starts to decay exponentially. It is
worth noting that for the three frequencies considered, the total
energy does not reveal any behavior similar to the diffusive
waves, in contrast to the previous theoretical
conjecture\cite{Sheng}. The theory predicts that the total energy
would follow the behavior of diffusive waves until when the sample
size is larger than the localization length. One explanation is
that in the present system, the scattering is too significant so
that the diffusive portion never dominates. A search for the
possible match between the simulation and theory in certain
conditions is still undergoing.

\begin{figure}[hbt]
\vspace{10pt} \epsfxsize=2.75in\epsffile{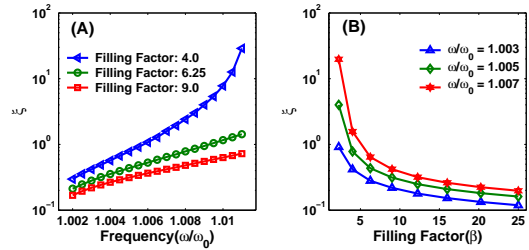}
\caption{Localization length as a function of (A) frequency and
(B) filling factor. The sample size is $8\times 8$ for (A).}
\label{fig9}
\end{figure}

In Fig.~\ref{fig9}, we plot the localization length versus
frequency and filling factor separately. It is shown that with the
fixed filling factor, referring to Fig.~\ref{fig9}(A), the
localization length increases with frequency within the frequency
regime considered. With a fixed frequency, the localization length
tends to decreases, meaning increasing localization effects, as
the filling factor increases.

\begin{figure}[hbt]
\vspace{10pt} \epsfxsize=2.75in\epsffile{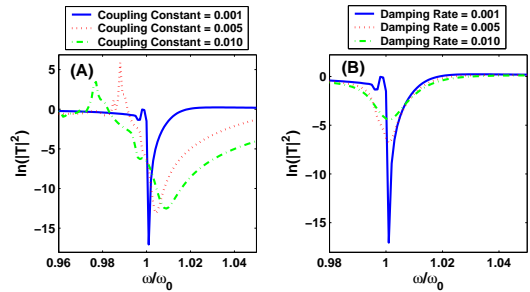}
\caption{Transmission versus frequency for various coupling
strengths and damping factors: (A) various coupling strengths with
damping rate 0.001; (B) various damping rates with coupling
constant 0.001. The sample size is $4\times 4$.} \label{fig10}
\end{figure}

Fig.~\ref{fig10} shows the transmission versus frequency for
various coupling constants and damping rates. From this figure, we
observe the following. (1) The increasing coupling strength leads
to a wider strong inhibition region, but shallower localization
valley. (2) When increasing the coupling strength, a prominent
resonance peak appears below the natural frequency $\omega_0$. And
the peak moves toward lower frequencies as the strength increases,
a feature also appears in the acoustic system\cite{Ye1}. (3)
Overall speaking, the increasing damping rate degrades the
localization level, and tends to abolish the resonance peak. Also
it seems to widen the strong localization region at the lower
frequency side.

\section{Summary}

In this article, the localization features in a simple
electromagnetic system are investigated in detail. Some general
properties of the localization phenomenon are elaborated. For
certain ranges of frequencies, strongly localized electromagnetic
waves have been observed in such a simple but realistic disordered
system. It is shown that the localization depends on a number of
parameters including frequency, filling factor, and damping rate.
The spatial behavior of the total, coherent and diffusive waves is
also explored, and found to comply with a physical intuitive
picture. A phase diagram characterizing localization is presented,
in agreement with previous investigations on other
systems\cite{Ye2}.

\section*{Acknowledgments}

This work is supported by the National Science Council of Republic
of China and Department of Physics at NCU.


\begin{thebibliography}{99}

\bibitem{Anderson} P. W. Anderson, Phys. Rev. {\bf 109}, 1492
(1958).

\bibitem{Lee} P. A. Lee and Ramakrishnan, Rev. Mod. Phys. {\bf
57}, 287 (1985).

\bibitem{Seis} R. Hennino, {\it et. al.}, Phys. Rev. Lett. {\bf 86},  3447 (2001).

\bibitem{ocean} G. I. Barenbatt, M. E. Vinogradov, and S. V.
Petrovskii, Oceanology, {\bf 35}, 202 (1995).

\bibitem{Laser} Z. Q. Zhang, Phys. Rev. B {\bf 52}, 7960 (1995);
S. Wiersma, M. P. van Albada, and A. Lagendijk, Nature, {\bf 373},
203 (1997); H. Cao, et. al. Phys. Rev. Lett. {\bf 82}, 2278
(1999); C. M. Soukoulis, X. Jiang, J. Y. Xu, and H. Cao, Phys.
Rev.  B {\bf 65}, 041103 (2002).

\bibitem{Thouless} D. J. Thouless, Phys. Rep. {\bf 13}, 93 (1974).

\bibitem{John} S. John, Phys. Today, {\bf 44}, 52 (1991).

\bibitem{Sheng} P. Sheng, {\it Introduction to Wave Scattering,
 Localization, and Mesoscopic Phenomena} (Academic Press, New York,
1995).

\bibitem{Lagen} A. Lagendijk and B. A. van Tiggelen, Phys. Rep.
{\bf 270}, 143 (1996).

\bibitem{Van} B. A. van Tiggelen, in {\it Diffuse Waves in Compex
Media} (Kluwer Academic Publisher, Netherlands, 1999).

\bibitem{Rossum} M. C. W. van Rossum and Th. M. Nieuwenhuizen,
Rev. Mod. Phys. {\it 71}, 313 (1999)

\bibitem{loc} M. Janssen, {\it Fluctuations and localization}
(World Scientific, Singapore, 2001); and references therein.

\bibitem{HW} C. H. Hodges and J. Woodhouse, J. Acoust. Soc. Am.
{\bf 74}, 894 (1983).

\bibitem{Kirk1} T. R. Kirkpatrick, Phys. Rev. B {\bf 31}, 5746 (1985).

\bibitem{He} S. He and J. D. Maynard, Phys. Rev. Lett. {\bf 57},
3171 (1986).

\bibitem{Condat} C. A. Condat, J. Acoust. Soc. Am. {\bf 83}, 441 (1988).

\bibitem{Sor} D. Sornette and B. Souillard, Europhys. Letts. {\bf 7}, 269
(1988).

\bibitem{Genack} A. Z. Genack and N. Garcia, Phys. Rev. Lett. {\bf 66}, 2064
(1991).

\bibitem{McCall} R. Dalichaouch, J. P. Armstrong, S. Schultz, P. M. Platzman
and S. L. McCall, Nature {\bf 354}, 53 (1991).

\bibitem{McCall2} S. L. McCall, P. M. Platzman, R. Dalichaouch, D.
Smith, and S. Schultz, Phys. Rev. Lett. {\bf 67}, 2017 (1991).

\bibitem{Marian} M. Rusek and A. Orlowski, Phys. Rev. E {\bf 51},
R2763 (1995).

\bibitem{Sigalas} M. M. Sigalas, C. M. Soukoulis, C.-T. Chan, and
D. Turner, Phys. Rev. B {\bf 53}, 8340 (1996).

\bibitem{AAA} A. A. Asatryan, et al. Phys. Rev. B {\bf 57}, 13535
(1998).

\bibitem{Ye1} Z. Ye and A. Alvarez, Phys. Rev. Lett. {\bf 80},
3503 (1998).

\bibitem{weak} Y. Kuga, A. Ishimaru, J. Opt. Soc. Am. A1, 831 (1984);
M. van Albada, A. Lagendijk, Phys. Rev. Lett. 55, 2692 (1985); P.
E. Wolf, G. Maret, Phys. Rev. Lett. 55, 2696 (1985); A. Tourin, et
al., Phys. Rev. Lett. 79, 3637 (1997); M. Torres, J. P. Adrados,
F. R. Montero de Espinosa, Nature {\bf 398}, 114 (1999).

\bibitem{Wiersma} D. S. Wiersma, P. Bartolini, A. Lagendijk, and R.
Roghini, Nature {\bf 390}, 671 (1997).

\bibitem{AAC} A. A. Chabanov, M. Stoytchev, and A. Z. Genack, Nature {\bf
404}, 850 (2000).

\bibitem{debate} F. Scheffold, R. Lenke, R. Tweer, and G. Maret,
Nature {\bf 398}, 206 (1999);  D. Wiersma, et al. Nature, {\bf
398}, 207 (1999).

\bibitem{Chen2002} Y.-Y. Chen and Z. Ye, Phys. Rev. E {\bf 65},
056612 (2002).

\bibitem{YLS} Z. Ye, S. Li, and X. Sun, Phys. Rev.
E {\bf 66}, 045602(R) (2002).

\bibitem{Erdogan} T. Erdogan, K. G. Sullivan, and D. G. Hall, J.
Opt. Soc. Am. B {\bf 10}, 391 (1993).

\bibitem{ccw} C.-C. Wang and Z. Ye, Phys. Stat. Sol. (a){\bf
174}, 527 (1999).

\bibitem{Zallen} R. Zallen, {\it The physics of amorphous solids},
(John Wiley and Son, New York, 1983).

\bibitem{Ye2} E. Hoskinson, and Z. Ye, Phys. Rev. Lett.
       {\bf 83}, 2734 (1999).

\bibitem{Gurtvitz} S. A. Gurtvitz, Phys. Rev. Lett. {\bf 85}, 812
(2000); and references therein.

\bibitem{3D} Z. Ye, H.-R. Hsu, E. Hoskinson, and A. Alvarez, Chin. J. Phys., {\bf 37}, 343
(1999).

\end{thebibliography}
\end{document}